\documentclass[letterpaper,english]{iopart}
\usepackage{iopams}
\usepackage[T1]{fontenc}
\usepackage[latin9]{inputenc}
\usepackage{color}
\usepackage{booktabs}
\usepackage{amssymb}
\usepackage{graphicx}
\usepackage[numbers]{natbib}

\providecommand{\tabularnewline}{\\}

\usepackage{babel}
\begin{document}
\title{Ionisation rate and Stark shift of a one-dimensional model of the
Hydrogen molecular ion}
\author{J. C. G. Henriques$^{1}$, Thomas G. Pedersen$^{2,3}$, N. M. R. Peres$^{1,4}$}
\address{$^{1}$Department and Centre of Physics, and QuantaLab, University
of Minho, Campus of Gualtar, 4710-057, Braga, Portugal}
\address{$^{2}$Department of Materials and Production, Aalborg University,
DK-9220 Aalborg {\O}st, Denmark}
\address{$^{3}$Center for Nanostructured Graphene (CNG), DK-9220 Aalborg {\O}st,
Denmark}
\address{$^{4}$International Iberian Nanotechnology Laboratory (INL), Av.
Mestre José Veiga, 4715-330, Braga, Portugal}
\begin{abstract}
In this paper we study the ionization rate and the Stark shift of
a one-dimensional model of the H$_{2}^{+}$ ion. The finding of these
two quantities is reduced to the solutions of a complex eigenvalue
problem. We solve this problem both numerically and analytically.
In the latter case we consider the regime of small external electrostatic
fields and small internuclear distances. We find an excellent agreement
between the ionization rate computed with the two approaches, even
when the approximate result is pushed beyond its expected validity.
The ionization rate is very sensitive to small changes of the external
electrostatic field, spanning many orders of magnitude for small changes
of the intensity of the external field. The dependence of the ionization
on the internuclear distance is also studied, as this has a direct
connection with experimental methods in molecular physics. It is shown
that for large distances the ionization rate saturates, which is a
direct consequence of the behavior of the energy eigenvalue with the
internuclear distance. The Stark shift is computed and from it we
extract the static polarizability of H$_{2}^{+}$ and compare our
results with those found by other authors using more sophisticated
methods. 
\end{abstract}
\maketitle

\section{Introduction}

The quantum study of the hydrogen molecular ion is almost as old as
quantum mechanics itself \citep{pauling1928application}. This molecular
ion is the simplest system we can think of in molecular physics. It
consists of a three body fermionic problem, where two of the fermions
are the identical protons and the third one is the electron binding
the two protons together. This system configures a three-body problem.
The wave functions, energy levels, and equilibrium distances are known
from numerical calculations \citep{chelkowski1992ionization} or from
variational approaches \citep{clark1969three,arista1973variational,frolov_2002,dunne2004simple}.
When acted upon by a static electric field the system can ionize following
two different channels \citep{nagaya2004laser,dietrich1992ionization}:
(i) the Coulomb explosion channel (or ionization channel), where the
reaction has the form $\mathrm{H}_{2}^{+}\rightarrow p+p+e$; (ii)
the dissociation channel, where the reaction form has the form $\mathrm{H}_{2}^{+}\rightarrow p+\mathrm{H}$.
These two channels have different probabilities to occur, depending
on whether we are considering an electrostatic field or a laser field,
from which a number of photons can be absorbed.

The motion of the three particles involved in the dynamics of the
$\mathrm{H}_{2}^{+}$ ion can be separated in two different types
of motion: (i) slow motion of the heavy protons; (ii) the fast motion
of the electron. As noted, solving the three-body problem considering
the motion of the three particles in the same footing is virtually
impossible from an analytical point of view. However, because of the
two very different time scales involved in the motion dynamics we
can separate the motion of the protons from the motion of the electron.
This approach is at the heart of the Born-Oppenheimer approximation.
Therefore, one usually considers a time interval where the protons'
motion is frozen and the electron dynamics is considered as the only
relevant process. 

Taking into account the Born-Oppenheimer approximation, analytical
expressions for the ground state wave function of $\mathrm{H}_{2}^{+}$
are obtained using an appropriate set of coordinates \citep{arista1973variational}.
The ionization of the ion requires adding an extra term to the Hamiltonian
that describing the interaction of the electron with the static electric
field. This apparently simple term, linear in position, immediately
renders the molecular problem intractable, even within the Born-Oppenheimer
approximation. This is a consequence of the known difficulty of solving
the Coulomb problem in the presence of an external electric field
\citep{Damburg_1976}. 

All the aforementioned difficulties can be overcome by studying simpler
models, which still contain the basic physical traits of the problem.
It is in this context that simpler one-dimensional models \citep{loudon2016one,loudon1959one,fernandez1985stark,blumel1987microwave,blumel2006analytical,fillion2012relativistic,dutta2008cartoon,foldy1976interesting,yao1992strong,nielsen1978simplified,postma1984polarizability,postma1985photoelectric,buchleitner1995spectral,dunne2004simple,lapidus1970one}
are usually proposed for studying the ionization of atoms and molecules
by static electric fields. In these types of models, the Coulomb interaction
is replaced by a short range potential, typically a Dirac $\delta$-function
potential, whose strength is adjusted to produce the observed ionization
energy of the atom or molecule in case (see procedure further down
in the text). In addition to this, the coupling to the external electrostatic
field is added via the dipole approximation. These models are much
simpler than the Coulomb problem, since the short range nature of
the potential guaranties that the effect of the external field can
be obtained exactly.

Many different problems related to this topic have been studied. As
mentioned above, approaches based on numerical and variational methods
have been implemented. Methods based on cylindrical \citep{chelkowski1992ionization},
elliptic \citep{arista1973variational} and parabolic \citep{bisgaard2004tunneling}
coordinates have been used. Ionization by static and laser fields
\citep{austin1979ionisation,yao1992strong,geltman1994short} has been
analyzed. In addition, the use of sum rules has been considered to
study the problem of a single $\delta-$function in Ref. \citep{belloni2008quantum}
and the problem of dissociation and Stark effect in non-rigid dipolar
molecules in Ref. \citep{pedersen2020hypergeometric}. The delta function
approach to the Coulomb potential in 1D has been used by many authors.
In Ref.\citep{austin1979ionisation} the problem of a $\delta-$function
potential in a strong field is treated numerically. In Ref. \citep{blumel2006analytical},
absolutely convergent periodic-orbit expansion techniques are used
to obtain an analytical solution for the compressed Hydrogen atom,
where the finite-range Coulomb potential is replaced by a zero-range
$\delta-$function potential. In Ref. \citep{lapidus1970one} a simple
model of a diatomic ion using $\delta-$functions is compared with
more complex methods. The $\delta-$function potential has also been
used to unveil the optical response of one-dimensional semiconductors,
that is the response to an electric field oscillating in time \citep{pedersen2015}.

The success of the $\delta-$function potential to describe the basic
physics of Coulomb explosion of simple molecules and ions has been
so impressive that we follow the same path in this paper and analyze
a different approach to the calculation of the ionization rate of
the $\mathrm{H}_{2}^{+}$ ion, which has not been considered before
in the literature to our best knowledge. 

The paper is organized as follows: in Sec. \ref{sec:Model} we introduce
the molecular-ion model, which has already been studied by many authors,
mainly to defined the problem and to fix the notation used throughout
the paper. In Sec. \ref{sec:Ionization-rate-and}, the ionization
rate and the Stark shift for our one-dimensional ion model are computed.
In Sec. \ref{sec:Summary} we give a summary of our results.

\section{Model\label{sec:Model}}

In this section, we introduce the model for the molecular ion and
define the different quantities used throughout the text. We start
solving the problem in the absence of an external static electric
field, thus computing the bound states of the system. When an external
electric field is present the bound states become quasi-bound states,
presenting an energy shift relative to the zero field case, as well
as a finite spectral width. This will be studied in the next section.

The Schrödinger equation, in atomic units, describing our one-dimensional
model of the H$_{2}^{+}$ ion reads
\begin{equation}
-\frac{d^{2}}{dx^{2}}\psi(x)-  W\left[\delta(x-a)+\delta(x+a)\right]\psi(x)
+  Fx\psi(x)=-\kappa^{2}\psi(x),
\end{equation}
where $W=2\mu Z$ (with $Z$ playing the role of
the nuclear charge), $F=2\mu\lambda$ (playing the role of the electric
force), and $\kappa^{2}=-2\mu E$, with $E$ the binding energy associated
with the wave function $\psi(x)$. Both $W$ and $F$ are positive
quantities, and $E<0$ when $F\rightarrow0$ but is otherwise a complex
number. he quantities $\mu$, $Z$, $\lambda$,
and $a$ refer to the reduced mass, potential strength or nuclear
charge, force field strength, and half the internuclear separation
(half the proton-proton distance), respectively. In the Born-Oppenheimer
approximation, where the electronic and nuclear motions are decoupled,
this equation can be seen as a one-dimensional model of a homo-nuclear
diatomic molecule, such as $\mathrm{H}_{2}^{+}$. In this particular
case, the two protons have an equilibrium distance given by $2a$,
and the single electron experiences two $\delta-$like potentials
associated with each of the two protons. This specific example will
be the object of our attention later in the text. In a more realistic
model, the single electron should interact with the two protons through
the Coulomb interaction. However, these two potentials, the $\delta-$function
potential and the Coulomb potential, share some common features, namely:
both diverge at the origin; the virial theorem is the same in both
cases \citep{foldy1976interesting}, and both respect the property
$xV'(x)=-V(x)$, with $V(x)=\delta(x)$ or $1/|x|$. These similarities,
combined with the simplicity the $\delta-$function potential brings
to the problem, make this a popular choice to model 1D systems. With
this said, one should keep in mind that the $\delta-$function potential
is still significantly different from the Coulomb potential, which
inevitably reduces the accuracy of the final results.Still, previous works report a fair agreement between
the $\delta-$function and the Coulomb potential, when both are used
to model the same system \citep{nielsen1978simplified}.
One way to overcome the differences between the $\delta-$function
potential and Coulomb potential is to ascribe a value for $W$ that
reproduces the ground state energy of the Coulomb potential (this
is the strategy we use later in the paper). 

We now focus on the case where the external electric field is absent,
that is $F=0$. In this case, when $x\neq\pm a$, the general solution
of the Schrödinger equation corresponds to a superposition of real
valued exponentials 
\begin{equation}
\psi(x)=Ae^{\kappa x}+Be^{-\kappa x},
\end{equation}
with $A$ and $B$ some coefficients determined by the constraints
of the problem, such as boundary conditions, continuity, and normalization.
The three different regions of our system will present particular
cases of this general solution. For $x<-a$ in order to have a finite
wave function, only the exponential with a positive argument can exist.
Following the same reasoning, for $x>a$ only the exponential with
a negative argument is allowed to be present. In the middle region,
$-a<x<a$, we use the symmetry of the system with respect to the origin
to state that the wave functions must have either even or odd parity,
that is, $\psi(x)=C\cosh(\kappa x)$ or $\psi(x)=C\sinh(\kappa x)$,
respectively, with $C$ some constant. These two cases have to be
studied separately. The introduction of a specific parity to our solution
simplifies the problem, since now we only need to study the boundary
conditions at either $a$ or $-a$.

Imposing the continuity of $\psi(x)$ at $x=a$, $\psi(a^{+})=\psi(a^{-})$,
as well as the discontinuity of its derivative, which is a direct
consequence of the $\delta-$functions in the Schrödinger equation,
$\psi'(a^{+})-\psi'(a^{-})=-W\psi(a)$, one obtains a system of two
equations with two unknowns. In order to guarantee the existence of
non-trivial solutions, we require the determinant of the system to
vanish. From here an implicit relation defining the binding energy
in the absence of the electric field is obtained as
\begin{eqnarray}
\tanh(\kappa_{0}^{\mathrm{even}}a) & =&\frac{W}{\kappa_{0}^{\mathrm{even}}}-1,\label{eq:k even}\\
\coth(\kappa_{0}^{\mathrm{odd}}a) & =&\frac{W}{\kappa_{0}^{\mathrm{odd}}}-1,\label{eq:k odd}
\end{eqnarray}
where $\kappa_{0}^{{\rm even/odd}}=\sqrt{-2\mu E_{0}^{{\rm even/odd}}},$
with $E_{0}^{{\rm even/odd}}$ the energy, in the absence of the field,
of the even/odd solution. Equations (\ref{eq:k even}) and (\ref{eq:k odd})
define two real eigenvalue problems. From the inspection of the two
previous relations we note that the first one, related to the even
case, always has a solution. However, the second one, associated with
the odd wave function, only has a solution when $aW>1$. This is easily
understood from the asymptotic behavior of the equation. When $\kappa_{0}^{\mathrm{odd}}a\rightarrow0$,
the left hand side diverges as $\coth(\kappa_{0}^{\mathrm{odd}}a)\sim1/\kappa_{0}^{\mathrm{odd}}a$
while the right hand side goes as $aW/\kappa_{0}^{\mathrm{odd}}a$.
In the opposite limit, when $\kappa_{0}^{\mathrm{odd}}a\rightarrow\infty$,
the left hand side approaches $1$, while the right hand side approaches
$-1$. In order to find a solution both sides of the equation have
to intercept at some specific value of $\kappa_{0}^{\mathrm{odd}}a$.
This is only possible if the right hand side is more divergent than
the left hand side when $\kappa_{0}^{\mathrm{odd}}a\rightarrow0$,
which is equivalent to saying that a solution only exists when the
condition $aW>1$ is fulfilled. Finally we observe that as $a$ increases
the energies of the odd and even states get closer to each other and
to the saturation value $-W^{2}/2\mu$. Saying it differently, the
bonding and anti-bonding molecular orbitals separate from each other
as the internuclear distance is decreased due to hybridization of
the two atomic orbitals.

\section{Ionization rate and Stark shift\label{sec:Ionization-rate-and}}

In this section, we draw inspiration from Ref. \citep{fernandez1985stark}
to extend the previous discussion to the case where the external electric
field is present. In the first part of this section we will obtain
an approximate analytical expression for the Stark-shift and the ionization
rate. Afterwards, we will focus on the specific case of the hydrogen
molecule ion $\mathrm{H}_{2}^{+}$ and study how the Stark-shift and
ionization rate depend on the parameters of the system. In
the end, we will compute the static polarizability and compare it
with previously published results, both theoretical and experimental.

In the presence of an external electric field, $F\neq0$, the Schrödinger
equation when $x\neq\pm a$ becomes
\begin{equation}
\frac{d^{2}}{dx^{2}}\psi(x)-F\left(x+\frac{\kappa^{2}}{F}\right)\psi(x)=0.
\end{equation}
With the change of variable $y=F^{1/3}(x+\kappa^{2}/F)$, the equation
can be cast in the form
\begin{equation}
\frac{d^{2}}{dy^{2}}\psi(y)-y\psi(y)=0,\label{eq:Airy Eq}
\end{equation}
which is nothing but the Airy differential equation, whose solution
is a superposition of the Airy functions $\mathrm{Ai}(y)$ and $\mathrm{Bi}(y)$
\citep{abramowitz1988handbook,vallee2004airy}. We can now proceed
in a similar way to what was done in the previous section, only this
time we will be working with Airy functions instead of real valued
exponentials, a clear sign of the change of behavior of the system
in the presence of an external electrostatic field. The main difficulty
relatively to what was done before is that, in the presence of the
external electrostatic field, we can no longer use the parity of the
wave function to simplify the problem. This implies that the boundary
conditions have to be evaluated at both $x=a$ and $x=-a$, transforming
the problem into a more involved one.

Let us now briefly go over the reasoning behind solving Eq. (\ref{eq:Airy Eq})
in the three different regions of the system. As was said before,
the solution of the Airy equation will be a superposition of Airy
functions, so, in the region $x<-a$, the wave function has the general
form
\begin{equation}
\psi(y)=A\mathrm{Ai}(y)+B\mathrm{Bi}(y),
\end{equation}
In the previous section, we were able to simplify the solution in
this region by looking at the limit $x\rightarrow-\infty$ and demanding
the wave function to be finite. In the present scenario, the same
limit proves to be useful, although the employed physical argument
differs slightly. In the presence of an electrostatic field, when
$x\rightarrow-\infty$, the wave function has to present a traveling
wave nature. From the appropriate series expansion of both Airy functions,
we see that $\mathrm{Ai}(y)$ and $\mathrm{Bi}(y)$ are proportional
to sines and cosines, respectively. We thus see that in order to recover
a traveling wave in the $x\rightarrow-\infty$ limit, the superposition
coefficients $A$ and $B$ have to be related as $A=iB$. In the opposite
region, where $x>a$, the solution is once more a superposition of
Airy functions that can be simplified upon analysis of the correct
limiting case. Studying the limit $x\rightarrow\infty$, we note that
the Airy function $\mathrm{Bi}(y)$ is proportional to the real valued
exponential $e^{y}$. Since this asymptotic behavior would lead to
a divergence of the wave function, it is clear that the coefficient
associated with $\mathrm{Bi}(y)$ must vanish. Finally, in the intermediate
region, $-a<x<a$, we once again find the solution to be a superposition
of Airy functions, only this time there are no symmetry or physical
arguments that allow us to relate the coefficients of the superposition,
and the general solution has to be used.

With the wave function found in the three regions of interest, apart
from some undetermined constants, we can formulate a complex eigenvalue
problem, from which the value of the energy of the quasi bound state
can be obtained. Applying the boundary conditions at $x=a$ and $x=-a$,
and requiring the determinant of the resulting system to vanish, we
find the following implicit relation defining the eigenenergy as

\begin{eqnarray}
\frac{F^{2/3}}{W\pi}&=&F^{1/3}\mathrm{Ai}(y_{a})\mathrm{Ci}(y_{a})+F^{1/3}\mathrm{Ai}(y_{-a})\mathrm{Ci}(y_{-a})
\nonumber\\
&+&W\pi\mathrm{Ai}^{2}(y_{a})\mathrm{Bi}(y_{-a})\mathrm{Ci}(y_{-a})-W\pi\mathrm{Ai}(y_{a})\mathrm{Ai}(y_{-a})\mathrm{Bi}(y_{a})\mathrm{Ci}(y_{-a}),\label{eq:Mastereq}
\end{eqnarray}

with $y_{\pm a}=F^{1/3}(\pm a+\kappa^{2}/F)$ and $\mathrm{Ci}(z)=\mathrm{Bi}(z)+i\mathrm{Ai}(z)$.
In order to arrive at this result it was useful to consider the identity
$\mathrm{Ai}(z)\mathrm{Bi}'(z)-\mathrm{Ai}'(z)\mathrm{Bi}(z)=1/\pi$,
which is nothing but the Wronskian of the two Airy solutions. Equation
(\ref{eq:Mastereq}) has a higher degree of complexity than the one
found for the problem of a single $\delta-$function \citep{fernandez1985stark}.
The increased difficulty arises from the necessity of evaluating the
boundary conditions at two different coordinates, which transforms
the eigenvalue problem from a $2\times2$ to a $4\times4$ matrix
problem. Although Eq. (\ref{eq:Mastereq}) can be solved exactly
numerically, an analytical solution is not readily available and is
highly desirable. In order to find one, we will look at the weak field
limit $F\ll1$ (also corresponding to $y_{\pm a}\gg1$), and follow
a perturbative approach. We now introduce the following series representation
of the Airy functions in the limit of interest \citep{abramowitz1988handbook,vallee2004airy}
\begin{eqnarray}
\mathrm{Ai}(z) && \sim\frac{1}{2\sqrt{\pi}}z^{-1/4}e^{-\zeta}\sum_{k=0}^{\infty}(-1)^{k}\zeta^{-k}c_{k},\quad\zeta=\frac{2}{3}z^{3/2}\label{eq:Ai series}\\
\mathrm{Bi}(z) && \sim\frac{1}{\sqrt{\pi}}z^{-1/4}e^{\zeta}\sum_{k=0}^{\infty}c_{k}\zeta^{-k},\label{eq:Bi series}
\end{eqnarray}
where
\begin{equation}
c_{0}  =1\quad c_{k}=\frac{\Gamma(3k+1/2)}{54^{k}k!\Gamma(k+1/2)},\quad k>0.
\end{equation}
To continue with the calculation we insert the series expansions
into Eq. (\ref{eq:Mastereq}). From the simultaneous analysis of
the series expansions and of Eq. (\ref{eq:Mastereq}), one clearly
sees that, even if only the zero\textendash th order term of each
series was considered, a compact solution to the eigenvalue equation
would still be elusive due to the pre-factors in Eqs. (\ref{eq:Ai series})
and (\ref{eq:Bi series}). With this difficulty in mind, we introduce
the first approximation in our procedure by substituting all the $y_{\pm a}$
that appear in the pre-factors by $y_{0}=\kappa^{2}/F^{2/3}$. Although
at first sight this may seem too crude an approximation, the fact
that we will be concerned with the weak field limit justifies this
approach, since we can expect $\kappa^{2}/F$ to be considerably larger
than $a$, and thus $y_{\pm a}\approx y_{0}$ for a small enough field
strength. Furthermore, we stress that this approximation is only applied
to the pre\textendash factors, and not to the exponentials, where
even a small change in the argument could still lead to a significant
difference in the final result. This approximation simplifies our
problem, since we were now able to isolate a single $\kappa$ (which
is directly linked to the energy) on one side of the equation. After
this is done, one approximation remains: all the $\kappa$ that appear
on the opposite side of the equation are substituted by their value
in the absence of the electrostatic field, $\kappa_{0}$, as given
in Eq. (\ref{eq:k even}). Considering only the lowest order terms
in the different sums, a simple compact expression for $\kappa$ is
reached, which reads

\begin{eqnarray}
\kappa\approx W&+& i\frac{W}{4}\left(1+\frac{W}{2\kappa_{0}}\right)e^{-\frac{4}{3F}\left(\kappa_{0}^{2}Fa\right)^{3/2}}\nonumber\\
&+&
 i\frac{W}{4}\left(1-\frac{W}{2\kappa_{0}}\right)We^{-\frac{4}{3F}\left(\kappa_{0}^{2}-Fa\right)^{3/2}},\label{eq:k approx}
\end{eqnarray}
from which the energy $E$ can be obtained if one recalls that $\kappa=\sqrt{-2\mu E}$.
The real part of $E$ gives the energy of the Stark shifted energy
level, while the imaginary part yields the ionization rate $\Gamma$
according to the relation $\Gamma=-2\mathrm{Im}E/\hbar$. A comment
on the quality of this approximate result is now in order. As we will
shortly see, this result produces an excellent description of the
ionization rate as a function of the field strength $\lambda$ and
as a function of the internuclear distance $2a$. However, the same
cannot be said regarding the real part of the energy. On the one hand,
the lowest order approximation, although acceptable for the ionization
rate, fails to capture any information regarding the dependence of
the Stark shift on the field (below we show how to improve the value
of real part of the energy); for that to happen, higher order field
terms must be introduced in the calculation. On the other hand, the
zero\textendash th order term, which can be seen from Eq. (\ref{eq:k approx})
to be $\kappa=W$, or $E=-W^{2}/2\mu$, does not correspond to the
exact energy of the bound state in the absence of the field. In fact,
this is the energy we would obtain from Eq. (\ref{eq:k even}) if
we took the limit $a\rightarrow0$. It is important to note that,
since the binding energy in the absence of the field is given by an
implicit relation, Eq. (\ref{eq:k even}), one should not expect to
find the correct zero\textendash th order term with the approach we
followed, contrary to what would happen if the problem of a single
atom was treated \citep{fernandez1985stark}, since in this case the
analytical expression for the energy of the ground state is known.
Indeed, only the limit $a\rightarrow0$ should appear, which is exactly
what we find. However, when higher order terms are
included, the Stark shift (to be discussed below) should be accurately
determined, which, as we will see, it is.

\begin{figure}[h]
\begin{centering}
\includegraphics[width=8.6cm]{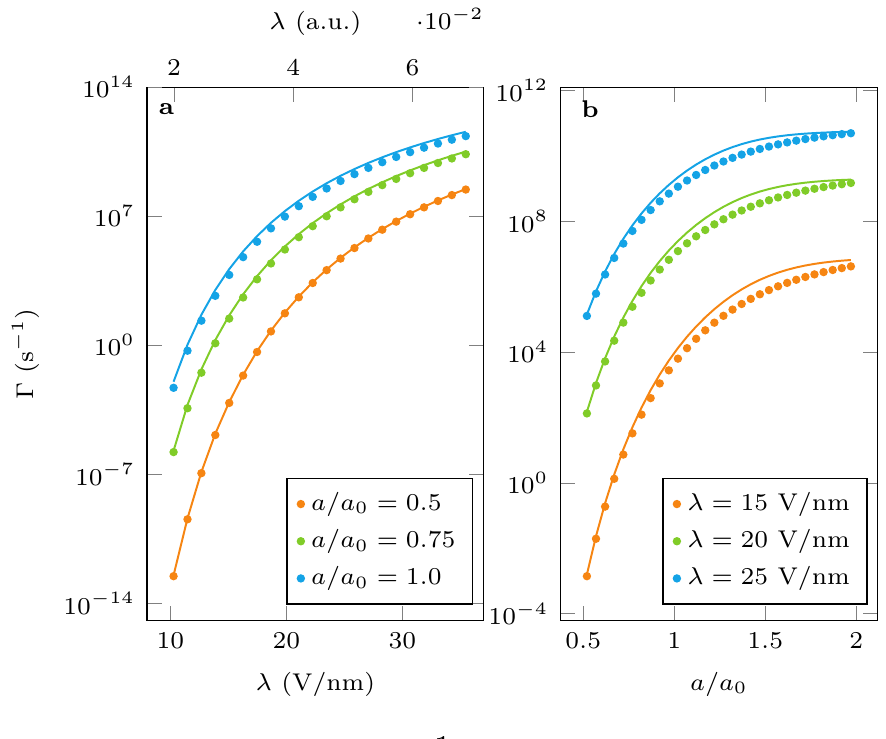}\caption{\label{fig:Ionization rate}Ionization rate of the one dimensional
H$_{2}^{+}$ ion. Panel \textbf{a} : Ionization rate $\Gamma$ as
a function of the field strength $\lambda$ for different values of
the internuclear separation. The solid lines represent the numerical
solution; the dots represent the approximate analytical
result. The agreement between the two data sets is clear and extends
across the different values of $a$. Panel \textbf{b}: Ionization
rate $\Gamma$ as a function of $a$ for three different electric
field strengths. Again, the solid lines represent the numerical
solution; the dots represent the approximate analytical
result. A good agreement between the numerical and analytical solutions
is seen. The ionization rate seems to saturate as $a$ increases,
similarly to the binding energy in the absence of the field, which
also saturates for large enough $a$. In both panels, we see that
the analytical solution (dots) slightly underestimates
the ionization rate. The values of Table \ref{tab:Parameters} were
used in both panels.}
\par\end{centering}
\end{figure}

Let us now explicitly put Eq. (\ref{eq:k approx}) to the test by
applying it to the case of the Hydrogen molecular ion. To do so, we
will consider the parameters given in Table \ref{tab:Parameters}.
\begin{table}[h]
\centering{}
\begin{tabular}{cccc}
\toprule 
$\mu$ ($m_{0}$) & $a$ ($a_{0}$) & $Z$ ($E_{h}a_{0}$) & $E_{0}^{\mathrm{even}}$ (eV)\tabularnewline
\midrule
\midrule 
1 & 1 &0.99 &$-16.45$\tabularnewline
\bottomrule
\end{tabular}\caption{\label{tab:Parameters}Parameters used to model
the Hydrogen molecular ion. The reduced mass $\mu$ was set equal
to $1$ bare electron mass ($m_{0})$, the equilibrium length 2$a$
was chosen to be equal to 2 Bohr radii ($a_{0})$, and the potential
strength was chosen as $0.99$ $E_{h}a_{0}$ (with $E_{h}$ corresponding
to the Hartree unit). This choice of $a$ and $Z$ reproduces the
total energy of H$_{2}^{+}$ (which we associate with the ground state
energy of the electron in the ion), which is known to be equal to
$-16.45$ eV, according to Ref. \citep{li2007totalenergy}. Also,
the equilibrium internuclear distance of $2a=2a_{0}$ is in agreement
with what is known for the ground state of H$_{2}^{+}$\citep{bell150,lai1977,li2007totalenergy}.}
\end{table}
In panel\textbf{ a }of Fig. \ref{fig:Ionization rate}, we depict
the ionization rate as a function of the field strength $\lambda$
from the numerical solution of Eq. (\ref{eq:Mastereq}) and from
the analytical expression given in Eq. (\ref{eq:k approx}), for different
values of $a$. There we observe an excellent agreement between the
approximate analytical expression and the exact numerical results,
even for values of $a\sim1$, which are beyond the expected validity
limit of our approximation. The ionization rate spans vast orders
of magnitude over a small range of field strengths, emphasizing the
extreme sensitivity of the former relative to the latter. Moreover,
we note that the ionization rate increases with increasing $a$ for
the same field strength. This is a result at the reduced binding energy.
For higher values of $a$ this increase becomes less noticeable. In
fact, we found reasonable results even for $a>1$, but due to its
similarity with the already depicted cases, we chose not to present
them. In the panel\textbf{ b }of the same figure, we study the dependence
of the ionization rate on the position $a$ of the protons, for three
different values of the field strength. This is a relevant behavior
to study due to its appearance in experimental works \citep{gibson1997strong,peng2004discrete}.
Once more, the agreement between the numerical and
analytical results is self\textendash evident. Furthermore, we observe
a saturation of the ionization rate as $a$ increases, which is consistent
with the behavior of the binding energy as a function of $a$, which
tends for a finite value , $E=-W^{2}/(2\mu)$, when $a\gg1$. Again
we note that the agreement between the data sets extends beyond the
expected validity region.

\begin{figure}[h]
\centering{}\includegraphics{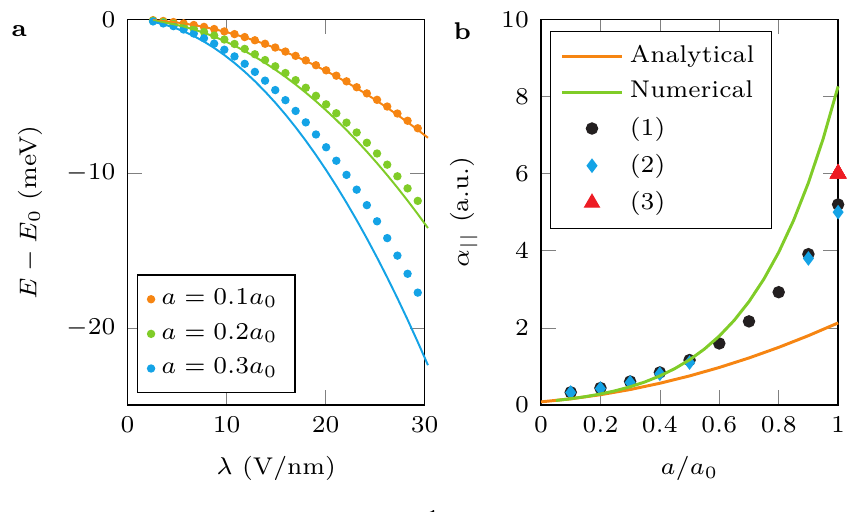}\caption{\label{fig:Stark-shift}Stark shift and polarizability of the one
dimensional H$_{2}^{+}$ ion. In panel \textbf{a} the Stark shift,
$E-E_{0},$ as function of the electrostatic field strength is depicted
for three different positions of the proton. The
solid lines represent the numerical solution; the dots represent the
approximate analytical result. A quadratic behavior with the field
strength is observed, since higher order field terms are less relevant
in the weak field regime we depict. As $a$ increases, the approximation
becomes worse, in agreement with the hypothesis of the expansion used.
In panel \textbf{b} the static polarizability of the one dimensional
H$_{2}^{+}$ ion computed from Eq. (\ref{eq:polarizability}) is compared
with that from Ref. \citep{McEachran1974_exp_pol} (1) and Ref. \citep{tsogbayar2009_exp_pol}
(2). The ``experimental'' value of $\alpha_{\parallel}$, estimated
via the relation $\alpha=(\alpha_{\parallel}+2\alpha_{\perp})/3$,
is represented by a red triangle (the value of $\alpha_{\perp}$ was
taken from Ref. \citep{tsogbayar2009_exp_pol}). The experimental
value of $\alpha$ was retrieved from Ref. \citep{sturrus1991_experimental,Jacobson1997}
(3). Also represented is the numerical calculation of the static polarizability
obtained from the numerically exact eigenvalues of Eq. (\ref{eq:Mastereq}). }
\end{figure}

Now, we turn to the Stark shift for the hydrogen molecular ion. As
we previously mentioned, Eq. (\ref{eq:k approx}) produces an excellent
description of the ionization rate, but fails to describe the correction
to the real part of the energy. We now wish to improve on this result
and obtain the field dependence of the energy shift. As was discussed
before, and is now stressed, the zero\textendash th order term in
the real part of the energy will never agree with the result in the
absence of the field, since the energy of the ion 
in zero field is given by the solution of a transcendent equation.
Because of this, we will be concerned with the Stark shift rather
than the shifted energy itself, that is, we will ignore the zero-th
order term, and only be interested in the terms containing information
about the electric field strength. Since now we
are only interested in the real part of the energy, we note that only
the real terms in Eq. (\ref{eq:Mastereq}) need to be considered
in further calculations (the imaginary ones are multiplied by vanishing
exponentials, which may be neglected in the study of the Stark shift).
When this simplification is performed, we introduce the series expansions
of the Airy functions in the eigenvalue equation and consider the
first two terms of the sums, an improvement regarding our first approach
where only the zero-th order was kept. Afterwards we expand the different
terms in the equation up to third and second order in $F$ and in
$a$, respectively, solve for $\kappa$ and once again expand up to
second order in both $F$ and $a$. Following this procedure, we obtain

\begin{eqnarray}
\kappa && \approx W\left(1-aW+2a^{2}W^{2}\right)\nonumber \\
 && +\left(\frac{5}{32W^{5}}+\frac{25a}{32W^{4}}+\frac{37a^{2}}{32W^{3}}\right)F^{2}.\label{eq:Stark}
\end{eqnarray}
From here the Stark shift is obtained using $\kappa\equiv\sqrt{-2\mu E}$
and subtracting the lowest order term. In Fig. \ref{fig:Stark-shift}
(panel \textbf{a}), we plot the Stark shift as a function of the field
strength for three different values of the internuclear distance.
A good agreement between the analytical and numerical results is seen
for small $a$. As this parameter increases the approximation becomes
worse, as expected, because we are moving outside its validity region.
We note that for the Stark shift the analytical results start to fail
for lower values of the internuclear distance than what was found
for the ionization rate, even after an improved expansion with higher
order terms considered. The reason behind this lies in the completely
different nature of these phenomena. The ionization rate is mainly
described by a real-valued exponential that is always present even
if only the lowest order is consider; this allows us to obtain excellent
results even with the simplest approach. The Stark shift, however,
is described by a power series, which requires a lot more care to
accurately capture the physics involved. In Fig. \ref{fig:Stark-shift}
(panel \textbf{b}), we depict the static polarisability,
$\alpha_{\parallel}$, defined via
\begin{eqnarray}
E^{(2)} && =-\frac{1}{2}\alpha_{\parallel}\lambda^{2}\nonumber \\
 && \approx-\frac{1}{2}\left(\frac{5}{64Z^{4}\mu^{3}}+\frac{5a}{8Z^{3}\mu^{2}}+\frac{11a^{2}}{8Z^{2}\mu}\right)\lambda^{2},\label{eq:polarizability}
\end{eqnarray}
where $E^{(2)}$ is the term associated with $\lambda^{2}$ in the
expansion of the real part of the energy, as function of the $a$
parameter. The first term of the previous expansion does not vanish
as $a\rightarrow0$ and corresponds to the polarizability of the ion
${\rm He}^{+}.$ Note that the polarizability decreases as the ``nuclear
charge'' $Z$ increases, as expected since the electron would be
more tightly bound.

Our analytical result is compared, in Fig. \ref{fig:Stark-shift},
with the data available (circles and diamonds) in the literature
using calculations considering the full three-dimensional nature of
the problem; the experimentally measured value for
the equilibrium internuclear distance is also depicted (triangle).
Given the simplicity of our approach, we believe we found a good agreement
between the two types of calculations. As expected, our result excels
in the region of small internuclear distance, and worsens as $a$
increases. At $a=1$, which gives the experimental polarizability
of the $\mathrm{H}_{2}^{+}$, our result is off by a factor of approximately
3, which, although not ideal, we consider to be good for the approximations
involved. The results of Eq. (\ref{eq:polarizability}) can be considerably
improved at large $a$ if the analytical exact polarizability of the
system is computed, which, in this case, is actually possible, although
not particularly illuminating. Still, in panel \textbf{\b
}  of Fig. \ref{fig:Stark-shift} we also represent
the polarizability computed numerically (green curve) from the exact
eigenvalues of Eq. (\ref{eq:Mastereq}). The agreement between the
analytical formula, given by Eq. (\ref{eq:polarizability}), and the
numerical result is excellent up to $a/a_{0}\lesssim0.4$. From hereon
the numerical curve departs from the result of Eq. (\ref{eq:polarizability})
growing faster with $a/a_{0}$. Given the simplicity of our model,
we consider that a good agreement was found between the analytical/numerical
polarizability and the data from other authors obtained from the three
dimensional study of the Hydrogen molecular ion.

\section{Summary\label{sec:Summary}}

In this work, the ionization rate and Stark shift of the hydrogen
molecule ion $\mathrm{H}_{2}^{+}$ in the presence of an external
static electric field were studied. From the Stark shift the static
polarizability was extracted. To model this system, we introduced
a one dimensional Schrödinger equation where the Coulomb interaction
between the two protons and the single electron was replaced by a
pair of $\delta-$functions.

In the first part of the text, we analyzed the system in the absence
of an electric field, studying the existence of its bound states and
their respective binding energies. Afterwards, the problem in the
presence of a finite electrostatic field was considered. Solving the
Schrödinger equation in the three different regions of the system,
we were able to write an eigenvalue problem, from which an implicit
relation defining the energy of the quasi-bound states was obtained.
This energy corresponds to a complex number whose real part gives
the central value of the binding energy, which is Stark shifted relative
to the zero field case, while the imaginary part is linked to the
line-width of the quasi-bound state or, saying it differently, to
the ionization rate. To find an approximate analytical expression
describing the energy we focused on the weak field and small internuclear
distance limit. Considering the leading order terms, we found an expression
for the ionization rate in excellent agreement with the exact numerical
results, even when the approximate solution was pushed beyond its
expected validity. We observed that the ionization rate was very sensitive
to the external electrostatic field strength, spanning many orders
of magnitude for small changes on the field's intensity.

The study of the dependence of the ionization rate on the internuclear
distance revealed that for large distances this quantity saturates,
which is a direct consequence of the dependence of the energy eigenvalue
on the same parameter. Although the leading order approximation produced
excellent results for the ionization rate, the same did not apply
to the real part of the energy. To improve this, higher order terms
had to be considered. Doing so allowed us to obtain results for the
Stark shift in good agreement with the numerical ones. However, the
zero\textendash th order term of the real part of the energy failed
to replicate the result in the absence of the electrostatic field.
This is easily understood because the exact eigenenergy comes from
the solution of a transcendental equation. Therefore, the zero-th
order term is the value of the energy for small $a$, which is to
be expected given the approximations made. Finally, the static polarizability
was computed and compared with data previously published in the literature
(for the three-dimensional ion) and a good agreement was found.

\ack

N.M.R.P. acknowledges support from the European Commission through
the project ``Graphene-Driven Revolutions in ICT and Beyond'' (Ref.
No. 881603 -{}- core 3), and the Portuguese Foundation for Science
and Technology (FCT) in the framework of the Strategic Financing UID/FIS/04650/2019.
JCGH acknowledges FCT for a scholarship in the context of the Summer
school ``Quantum Matter: Materials and Concepts''.



\end{document}